\documentclass[prl,reprint]{revtex4-1}
\pdfoutput=1
\usepackage{graphicx}
\usepackage{bm}
\usepackage{hyperref}

\begin{document}

\title{Anomalous robustness of the $\nu=5/2$ fractional quantum Hall state near a sharp phase boundary}
\date{today}

\author{Yang Liu}
\affiliation{Department of Electrical Engineering,
Princeton University, Princeton, New Jersey 08544}
\author{D.\ Kamburov}
\affiliation{Department of Electrical Engineering,
Princeton University, Princeton, New Jersey 08544}
\author{M.\ Shayegan}
\affiliation{Department of Electrical Engineering,
Princeton University, Princeton, New Jersey 08544}
\author{L.N.\ Pfeiffer}
\affiliation{Department of Electrical Engineering,
Princeton University, Princeton, New Jersey 08544}
\author{K.W.\ West}
\affiliation{Department of Electrical Engineering,
Princeton University, Princeton, New Jersey 08544}
\author{K.W.\ Baldwin}
\affiliation{Department of Electrical Engineering,
Princeton University, Princeton, New Jersey 08544}

\date{\today}

\begin{abstract}

  We report magneto-transport measurements in wide GaAs quantum wells
  with tunable density to probe the stability of the fractional
  quantum Hall effect at filling factor $\nu = $ 5/2 in the vicinity
  of the crossing between Landau levels (LLs) belonging to the
  different (symmetric and antisymmetric) electric subbands. When the
  Fermi energy ($E_F$) lies in the excited-state LL of the symmetric
  subband, the 5/2 quantum Hall state is surprisingly stable and gets
  even stronger near this crossing, and then suddenly disappears and
  turns into a metallic state once $E_F$ moves to the ground-state LL
  of the antisymmetric subband. The sharpness of this disappearance
  suggests a first-order transition.

\end{abstract}


\maketitle

There is tremendous interest currently in the origin and properties of
the fractional quantum Hall state (FQHS) at the even-denominator
Landau level (LL) filling factor $\nu = $ 5/2
\cite{Willett.PRL.1987}. This interest partly stems from the
expectation that the quasi-particle excitations of the 5/2 FQHS might
obey non-Abelian statistics \cite{Moore.Nuc.Phy.1991} and be useful
for topological quantum computing \cite{Nayak.Rev.Mod.Phys.2008}. The
stability and robustness of the 5/2 state, and its sensitivity to the
parameters of the two-dimensional electron system (2DES) in which it
is formed are therefore of paramount importance and have been studied
recently both experimentally \cite{Pan.PRL.1999, Pan.PRB.2008,
  Dean.PRL.2008, *Dean.PRL.2008.101, Choi.PRB.2008, Nuebler.PRB.2010,
  Shabani.PRL.2010, Xia.PRL.2010, Kumar.PRL.2010, Pan.PRL.2011} and
theoretically \cite{Rezayi.PRL.2000, Peterson.PRL.2008,
  *Peterson.PRB.2008, Papic.PRB.2009, Wojs.PRL.2010}.

Ordinarily, the 5/2 FQHS is seen in very low disorder 2DESs when the
Fermi energy ($E_F$) lies in the spin-up, excited-state ($N=1$), LL of
the ground-state (symmetric, S) electric subband, namely in the
S1$\uparrow$ level (see Fig. 1). It has been theoretically proposed
that a non-Abelian (Pfaffian) $\nu = $ 5/2 FQHS should be favored in a
"thick" electron system confined to a relatively wide quantum well
(QW) \cite{Rezayi.PRL.2000, Peterson.PRL.2008, Papic.PRB.2009,
  Wojs.PRL.2010}. But in a realistic, experimentally achievable wide
QW system, the electrons can occupy the second (antisymmetric, A)
electric subband. It was demonstrated very recently that, if the
subband energy spacing ($\Delta$) is smaller than the cyclotron energy
$\hbar\omega_c$, so that $E_F$ at $\nu = 5/2$ lies in the
\emph{ground-state} ($N = 0$) LL of the antisymmetric subband (i.e.,
in the A0$\uparrow$ level; see Fig. 1), then the $\nu=5/2$ FQHS is
destroyed and instead the standard, odd-denominator FQHSs
characteristic of the $N = 0$ LLs are seen
\cite{Shabani.PRL.2010,Liu.cond.mat.2011}. These observations imply
that the node in the \emph{in-plane} wave-function is crucial for the
stability of the 5/2 FQHS.

Here we examine the stability of the $\nu=5/2$ FQHS in relatively wide
GaAs QWs in the vicinity of the crossing (at $E_F$) between the
S1$\uparrow$ and the A0$\uparrow$ LLs (Fig. 1). We find that, when
$E_F$ lies in the S1$\uparrow$ LL, the 5/2 state is remarkably robust
and gets even stronger as the A0$\uparrow$ LL is brought to within
$\sim$ 1 K of the S1$\uparrow$ LL. As the crossing is reached and
$E_F$ moves into the A0$\uparrow$ LL, the $\nu = 5/2$ state abruptly
disappears.

\begin{figure}
\includegraphics[width=.48\textwidth]{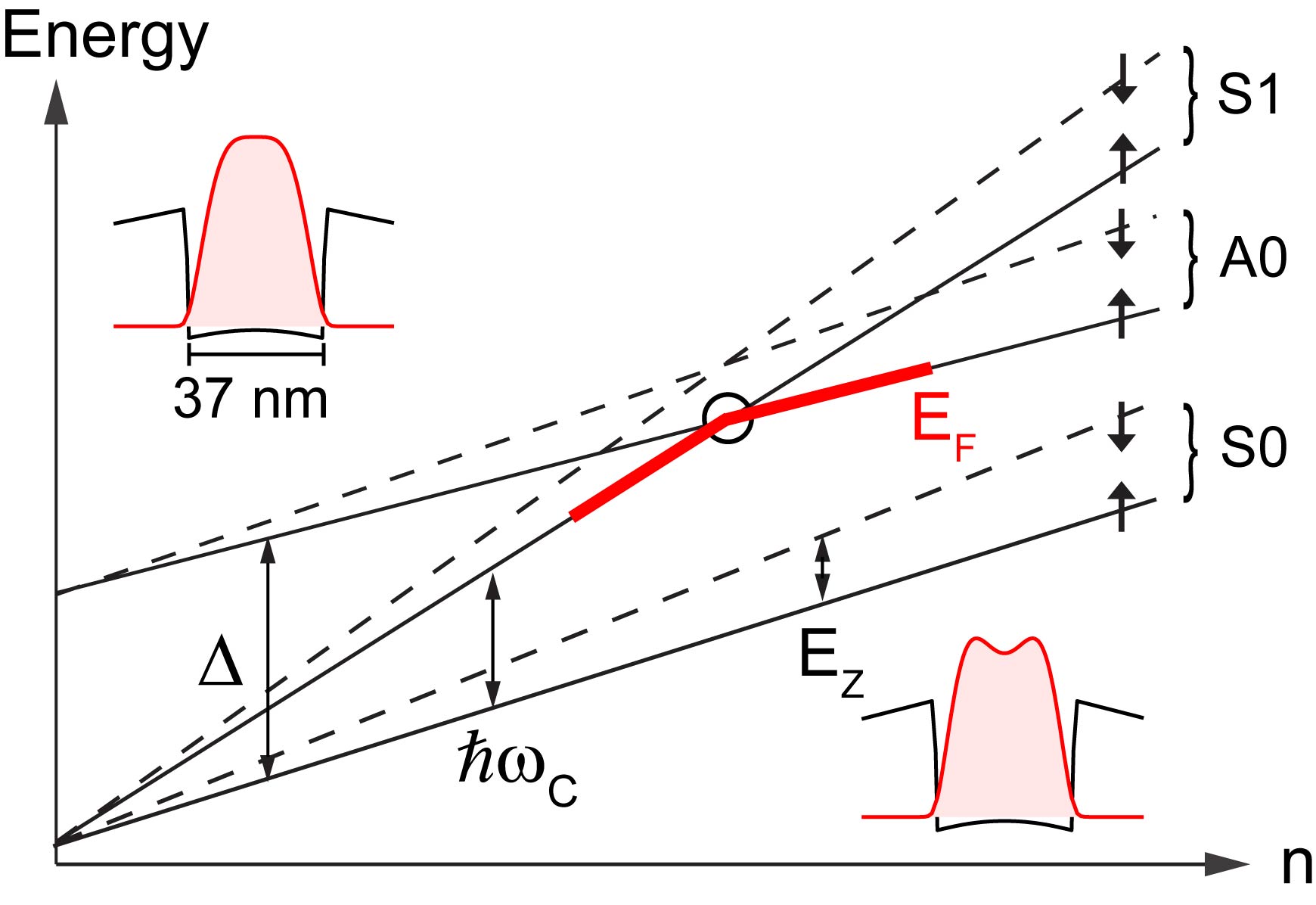}
\caption{\label{fig:cartoon}(color online) Schematic LL diagram for
  the symmetric (S) and antisymmetric (A) electric subbands as a
  function of increasing density $n$. The index 0 or 1 following S and
  A is the LL quantum number ($N$), and the up- ($\uparrow$) and
  down-spin ($\downarrow$) levels are represented by solid and dashed
  lines. The relevant energies are the subband separation ($\Delta$),
  the cyclotron energy ($\hbar\omega_c$), and the Zeeman energy
  ($E_Z$). As we increase $n$ while keeping the QW balanced,
  $\hbar\omega_c$ increases and $\Delta$ decreases. The S1$\uparrow$
  level crosses the A0$\uparrow$ level when
  $\hbar\omega_c=\Delta$. The Fermi energy (red line) moves from
  S1$\uparrow$ to A0$\uparrow$ at the crossing (marked by a
  circle). In our work, we study the evolution of the FQHSs near
  $\nu=5/2$ at this crossing. The upper left and lower right insets
  show the self-consistently calculated electron charge distributions
  (red curves) and potentials (black curves) at zero magnetic field
  for a 37-nm-wide QW, with densities of 2.09 and 2.48 $\times
  10^{11}$ cm${^{-2}}$, respectively.}
\end{figure}

\begin{figure}
\includegraphics[width=.48\textwidth]{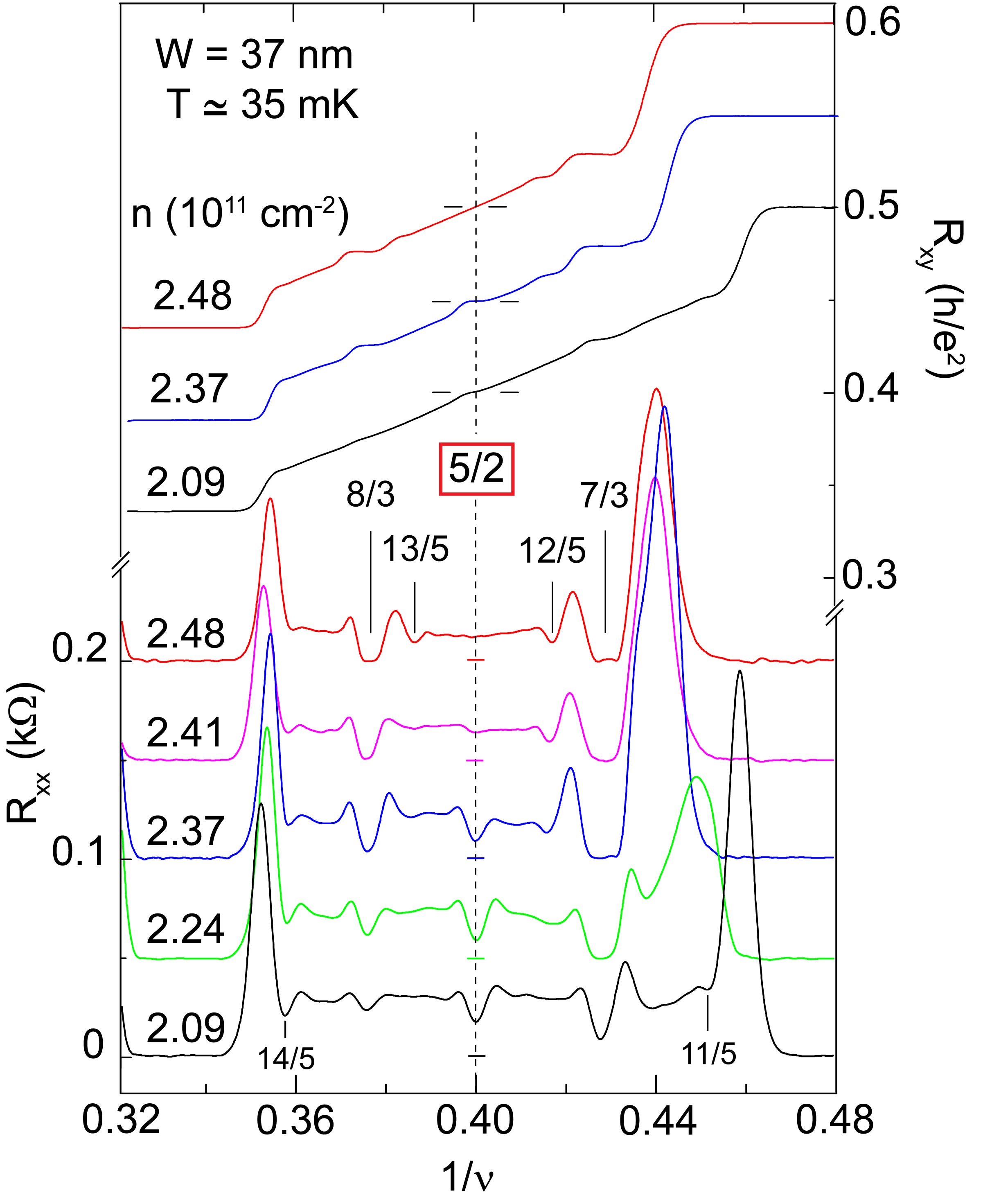}
\caption{\label{fig:waterfall} (color online) Waterfall plots of
  ${R_{xx}}$ and $R_{xy}$ magneto-resistances
  showing the evolution of FQHSs for the 37-nm-wide GaAs QW as the
  density is changed from $2.09$ to $2.48\times 10^{11}$
  cm${^{-2}}$. Except for the lowest density $R_{xx}$ and $R_{xy}$
  traces, for clarity each $R_{xx}$ trace is shifted vertically by 50
  $\Omega$, and each $R_{xy}$ trace by 0.05 $h/e^2$. }
\end{figure}

\begin{figure}[htb]
\includegraphics[width=.48\textwidth]{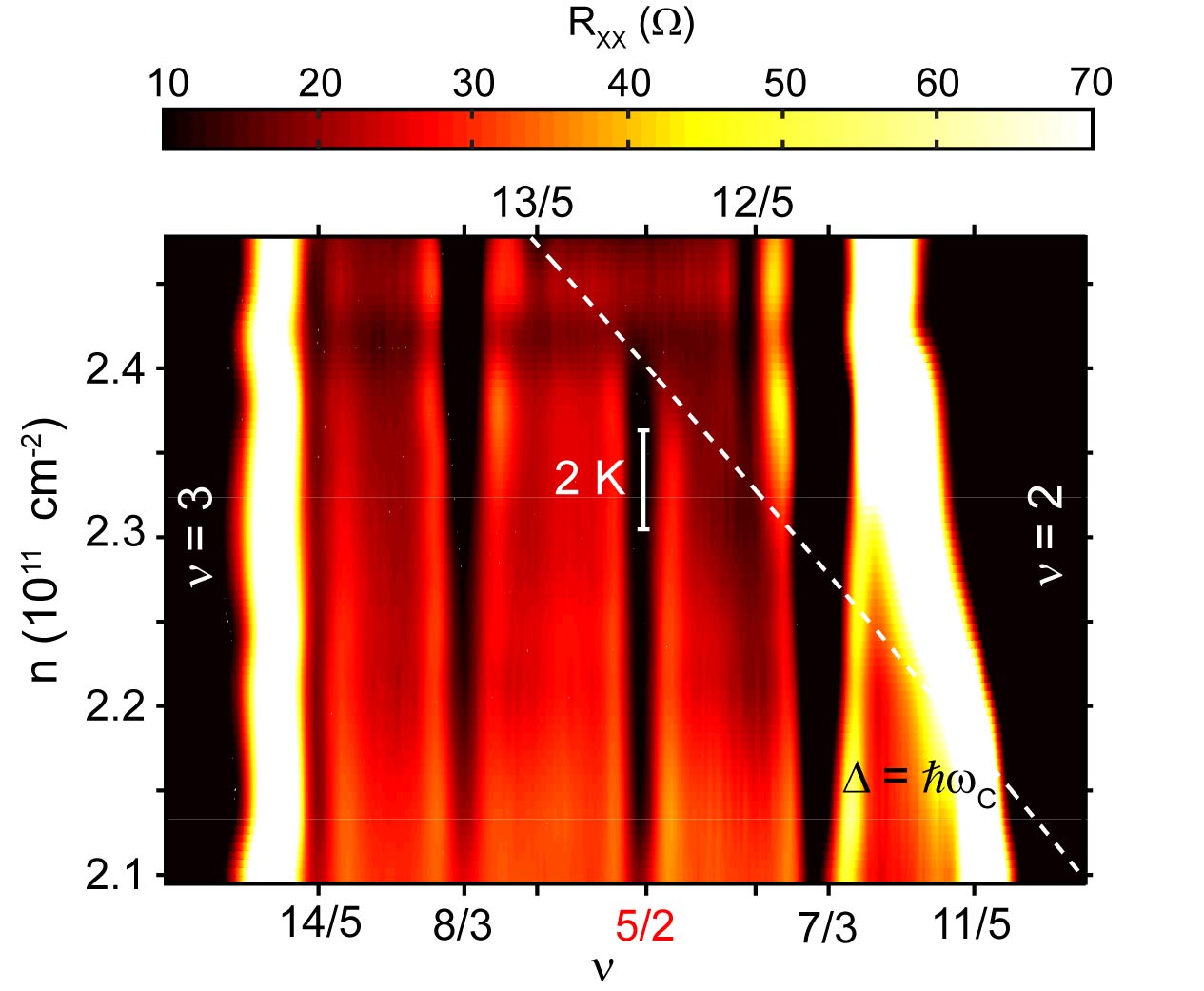}%
\caption{\label{fig:colorful} (color online) A color-scale plot of
  $R_{xx}$ for the 37-nm-wide QW demonstrating the evolution of the
  FQHSs as the density is increased from $n=2.09$ to $2.48\times
  10^{11}$ cm$^{-2}$. The bright regions correspond to large $R_{xx}$
  values, and the dark regions to small $R_{xx}$ where the quantum
  Hall states are observed. The white dashed line denotes the
  condition $\Delta = \hbar\omega_c$. Below (above) this line we
  expect $E_F$ to lie in the S1$\uparrow$ (A0$\uparrow$) level. The
  vertical bar provides an energy scale for the separation between the
  S1$\uparrow$ and A0$\uparrow$ levels at $\nu=5/2$. }
\end{figure}

\begin{figure}
\includegraphics[width=.48\textwidth]{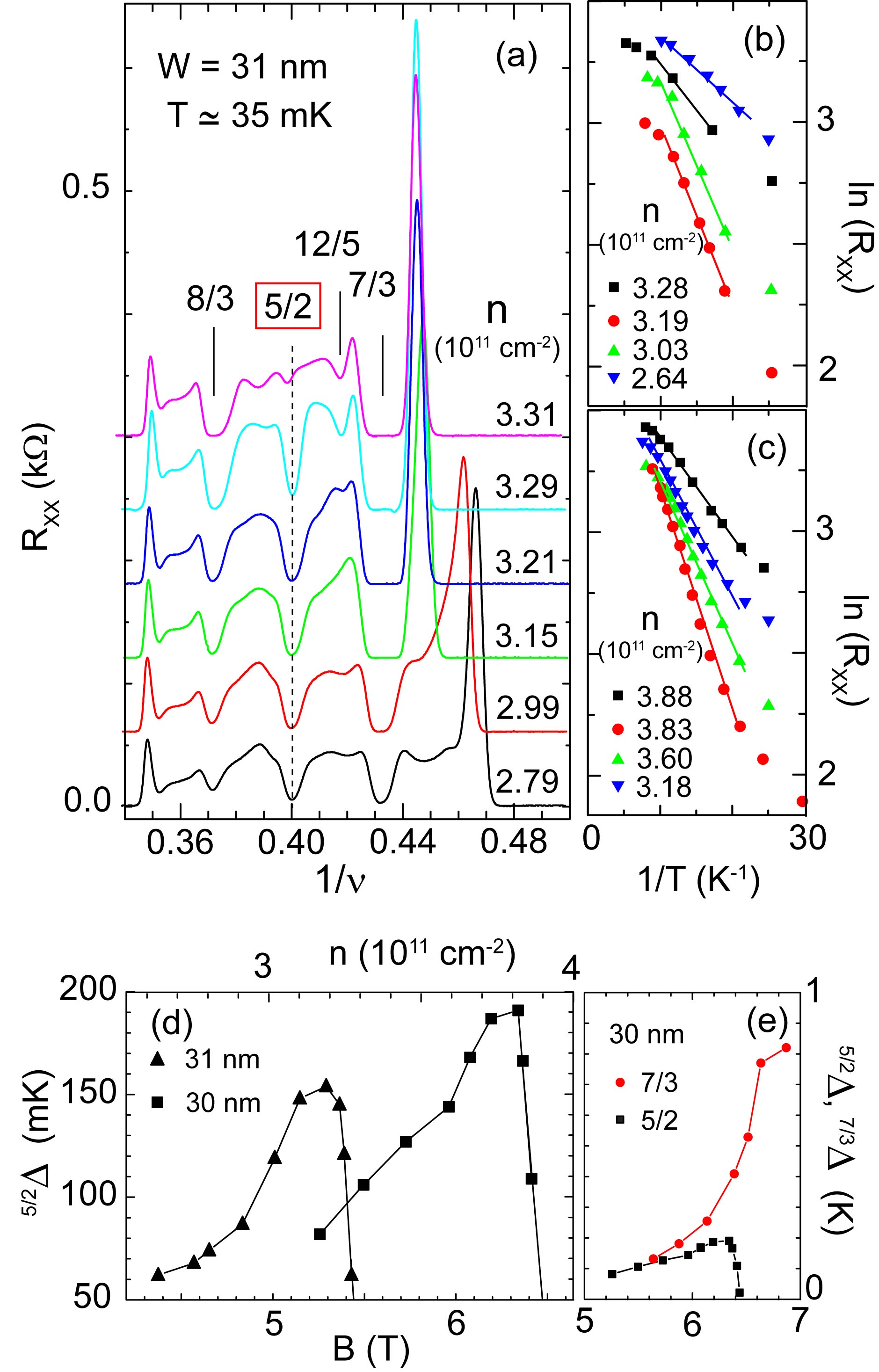}
\caption{\label{fig:waterfall2}(color online) (a) Waterfall plot of
  ${R_{xx}}$ vs. $1/\nu$ for the 31-nm-wide GaAs QW as $n$ is changed
  from $2.79$ to $3.31\times 10^{11}$ cm${^{-2}}$. The traces are
  shifted vertically (by 60 $\Omega$). (b) and (c) Arrhenius plots of
  $R_{xx}$ at $\nu = 5/2$ vs. inverse temperature for the 31- and
  30-nm-wide QWs at the indicated densities. Data are shifted
  vertically for clarity. (d) Measured energy gap for the $\nu=5/2$
  FQHS in both samples as a function of magnetic field or density. (e)
  Measured energy gaps for the $\nu=5/2$ and 7/3 FQHSs in the
  30-nm-wide QW.}
\end{figure}

Our samples were grown by molecular beam epitaxy, and each consist of
a wide GaAs QW bounded on each side by undoped
Al$_{0.24}$Ga$_{0.76}$As spacer layers and Si $\delta$-doped
layers. We report here data for three samples, with QW widths $W=$ 37,
31 and 30 nm, and densities of $n\simeq$ 2.5, 3.3 and 3.8 $\times
10^{11}$ cm$^{-2}$, respectively. The widths and the densities of
these samples were carefully designed so that, for each sample, its
$\Delta$ is close to $\hbar\omega_c$ at the magnetic field position of
$\nu=5/2$.  This enables us to make the S1$\uparrow$ and A0$\uparrow$
levels cross at $E_F$ by slightly tuning the density (Fig. 1), as we
describe below. The low-temperature ($T = 0.3$ K) mobilities of our
samples are $\mu \simeq$ 950, 480 and 670 m$^2$/Vs,
respectively. These are about a factor of three to four smaller than
the mobilities for 2DESs in single-subband QW samples grown in the
same molecular beam epitaxy chamber; we believe it is the occupancy of
the second electric subband that reduces the mobility in the samples
studied here.

Each of our samples has an evaporated Ti/Au front-gate and an In
back-gate. We carefully control $n$ and the charge distribution
symmetry in the QW by applying voltage biases to these gates
\cite{Shabani.PRL.2010,Suen.PRL.1994,Shabani.PRL.2009,
  Liu.cond.mat.2011}. For each $n$, we measure the occupied subband
electron densities from the Fourier transforms of the low-field ($B\le
0.5$ T) Shubnikov-de Haas oscillations. These Fourier transforms
exhibit two peaks whose frequencies, multiplied by $2e/h$, give the
subband densities, $n_S$ and $n_A$; see, e.g., Fig.~1 in
Ref. \cite{Shabani.PRL.2009}. The difference between these densities
directly gives the subband separation
$\Delta=\frac{\pi\hbar^2}{m^{*}}(n_S-n_A)$, where $m^{*} = 0.067m_e$
is the GaAs electron effective mass. All the data reported here were
taken by adjusting the front- and back-gate biases so that the total
charge distribution is symmetric. We add that our measured $\Delta$
agree well with the results of calculations that solve the Poisson and
Schroedinger equations to obtain the potential energy and the charge
distribution self-consistently.

Figure \ref{fig:waterfall} shows a series of longitudinal ($R_{xx}$)
and Hall ($R_{xy}$) magneto-resistance traces in the filling range $2
< \nu < 3$ for the 37-nm-wide QW sample, taken at different densities
ranging from 2.09 to $2.48\times 10^{11} $ cm$^{-2}$. As $n$ is
increased in this range, $\Delta$ decreases from 83 to 79 K while
$\hbar\omega_c$ at $\nu=5/2$ increases from 69 to 82 K, so we expect a
crossing of the S1$\uparrow$ and A0$\uparrow$ levels. This crossing
manifests itself in a remarkable evolution of the FQHSs as seen in
Fig. 2. At the lowest density, $R_{xx}$ shows reasonably
well-developed minima at $\nu$ = 5/2, 7/3, and 8/3, as well as weak
minima at 11/5 and 14/5. These minima are characteristic of the FQHSs
observed in high-quality, standard (single-subband) GaAs 2DESs, when
$E_F$ lies in the S1$\uparrow$ LL (see, e.g., Fig. 1(a) of
Ref. \cite{Shabani.PRL.2010}). At the highest $n$, the $R_{xx}$ minima
at $\nu = 5/2$, 11/5 and 14/5 \footnote{At yet higher $n$, which we
  cannot achieve in this sample, we expect the $\nu=$ 14/5 $R_{xx}$
  minimum to completely disappear once $E_F$ lies in the A0 level at
  $\nu = $ 14/5 (see, e.g., Fig. 1(b) in
  Ref. \onlinecite{Shabani.PRL.2010}).} have disappeared and instead
there are fully developed FQHSs at $\nu = $ 7/3 and 8/3 as well as
developing minima at 12/5 and 13/5 \footnote{A very weak $\nu=$ 12/5
  FQHS is observed in ultra-clean, single-subband samples when $E_F$
  lies in the S1 level at very low temperatures [J. S. Xia $et$ $al.$,
  Phys. Rev. Lett. {\bf93}, 176809 (2004)]. We do not see such a FQHS
  in our samples because of their lower mobilities and the higher
  temperature of our experiments.}. All these features are
characteristic of FQHSs when $E_F$ is in the A0$\uparrow$ LL
\cite{Shabani.PRL.2010,Liu.cond.mat.2011}.

To better highlight the evolution of the FQHSs observed in Fig. 2, in
Fig. 3 we show an interpolated, color-scale plot of $R_{xx}$ as a
function of filling and density. Both Figs. 2 and 3 show that as $n$
is increased, the evolution of the FQHSs takes place from high-field
(low $\nu$) to low-field (high $\nu$). The weak $R_{xx}$ minimum at
$\nu = $ 11/5 observed at the lowest $n$, e.g., disappears quickly as
$n$ is raised and is followed by a strengthening of the 7/3 (and then
the 12/5) FQHS at higher $n$. Then comes the disappearance of the 5/2
FQHS, and eventually the strengthening of the 8/3 (and 13/5) FQHSs and
weakening of the 14/5 $R_{xx}$ minimum at the highest $n$. Such
evolution is of course expected: Since the S1$\uparrow$ level crosses
the A0$\uparrow$ LL when $\Delta = \hbar\omega_c\propto
\frac{n}{\nu}$, we expect the crossing to occur at progressively
higher $\nu$ as $n$, and consequently $\hbar\omega_c$ at a given
$\nu$, increase. To assess the position of the expected crossing
quantitatively, in Fig. 3 we have included a dashed curve, marked
$\Delta = \hbar\omega_c$. The value of $\Delta$ for this line is based
on our measured $\Delta$ from low-field Shubnikov-de Haas oscillations
which agree with the results of our self-consistent calculations for a
37-nm-wide QW. While we cannot rule out the possibility that $\Delta$
is re-normalized at magnetic fields in the 2 $< \nu <$ 3 range, it
appears that the dashed line corresponds to the position of the LL
crossing accurately: the $\nu=12/5$ and 13/5 FQHSs, which are characteristic
of $E_F$ being in the A0$\uparrow$ level \cite{Note2}, are seen above
the dashed line, and the $\nu=5/2$ FQHS is seen only below this line when
$E_F$ lies in the S1$\uparrow$ LL. Interestingly, the $\nu=7/3$ FQHS
is observed on both sides of the dashed line and becomes stronger
monotonically as $n$ is raised.

Having established the crossing of the S1$\uparrow$ and A0$\uparrow$
LLs in Figs. 2 and 3, we now focus on our main finding, namely the
stability of the $\nu=5/2$ FQHS in the vicinity of this crossing. The
data of Figs. 2 and 3 indicate that as $n$ is raised, the 5/2 FQHS
initially becomes stronger. This strengthening is seen from the
deepening of the $R_{xx}$ minima, and particularly from the very well
developed $R_{xy}$ plateau at $n=2.37\times 10^{11}$ cm$^{-2}$
(compared, e.g., to the plateau for $n=2.09\times 10^{11}$ cm$^{-2}$,
see Fig. 2). We will return to this intriguing observation later in
the paper. Even more striking, however, is that \emph{the $\nu=5/2$
  FQHS, which is most robust at $n=2.37\times 10^{11}$ cm$^{-2}$,
  suddenly disappears when the density is increased by less than 2$\%$
  to $n=2.41\times 10^{11}$ cm$^{-2}$}.

Data for the narrower QW samples, presented in Fig. 4, verify the
above observations qualitatively. Moreover, they allow us to
quantitatively assess, through energy gap measurement, the robustness
of the $\nu=5/2$ FQHS near the LL crossing and the sharpness of its
disappearance. The $R_{xx}$ traces shown in Fig. 4(a) corroborate the
data of Fig. 2. A very similar evolution of the FQHSs is seen,
including a sudden disappearance of the 5/2 state at high $n$. Note
that $\hbar\omega_c$ at which the 5/2 FQHS disappears in Fig. 4(a) is
equal to 109 K, very close to the value of $\Delta\simeq$ 112 K for
this 31-nm-wide QW at $n=3.31\times 10^{11}$ cm$^{-2}$. From the
temperature dependence of the 5/2 $R_{xx}$ minimum (Fig. 4(b)), we are
also able to deduce an energy gap ($^{5/2}\Delta$) for the $\nu=$ 5/2
FQHS. The measured gap, shown in Fig. 4(d) as a function of the
magnetic field position of $\nu=$ 5/2, exhibits a behavior consistent
with the conclusions gleaned qualitatively from the $R_{xx}$ traces of
Figs. 2 and 4(a): $^{5/2}\Delta$ increases as $n$ is raised and then
suddenly decreases. Note in Fig. 4(d) that $^{5/2}\Delta$ collapses
from its maximum value when $n$ is increased by less than 3$\%$. The
sharpness of the collapse suggests that the ground state of the 2DES
makes a first-order transition from a FQHS to a metallic state as
$E_F$ moves from the S1$\uparrow$ to the A0$\uparrow$ level.


A remarkable feature of the data in Figs. 2-4 is that the $\nu=5/2$
FQHS becomes stronger with increasing $n$ before it collapses. This is
clearly evident in the plot of $^{5/2}\Delta$ vs. $B$ in Fig. 4(d). A
qualitatively similar increase of $^{5/2}\Delta$ with $n$ was seen
recently in 2DESs where only one electric subband was occupied
\cite{Nuebler.PRB.2010}, and was attributed to the enhancement of the
Coulomb energy and the screening of the disorder potential with
increasing $n$. It is possible that our data can be explained in a
similar fashion. However, the relatively steep rise of $^{5/2}\Delta$,
especially right before the collapse, is puzzling.

We have repeated the gap measurements for a slightly narrower
(30-nm-wide) QW and the data, shown in Figs. 4(c) and 4(d),
qualitatively confirm this anomalous behavior: $^{5/2}\Delta$
increases steeply with increasing $n$ and then suddenly drops once the
density exceeds $3.83\times 10^{11}$ cm$^{-2}$. Note that the higher
$n$, and therefore larger $\hbar\omega_c$, at which $^{5/2}\Delta$
collapses in the narrowest QW sample are consistent with its larger
subband separation. For this QW, at $n=3.88\times 10^{11}$ cm$^{-2}$,
we have $\Delta \simeq 130$ K, very close to $\hbar\omega_c=128$ K. A
noteworthy observation in Fig. 4(d) data is that, at a common density
of $n=3.2\times 10^{11}$ cm$^{-2}$, $^{5/2}\Delta$ for the wider (31
nm) QW is nearly twice larger than $^{5/2}\Delta$ for the narrower (30
nm) QW. The observation of a larger gap for a wider QW, which was also
reported in Ref.~\onlinecite{Xia.PRL.2010}, appears to be consistent
with the theoretical expectation that a Pfaffian $\nu=5/2$ FQHS should
be favored in a 2DES with larger electron layer thickness
\cite{Peterson.PRL.2008}. However, according to the available
calculations, while for thicker electron layers the overlap between
the numerically calculated wavefunction and the Pfaffian state is
enhanced, the energy gap is in fact reduced
\cite{Peterson.PRL.2008}. We conclude that the much larger gap
observed in Fig. 4(d) at $n=3.2\times 10^{11}$ cm$^{-2}$ for the
31-nm-wide QW sample compared to the 30-nm-wide sample is related to
the anomalous, steep rise of the gap before the LL crossing occurs.

In Fig. 4(e), we also show the energy gap of the $\nu=$ 7/3 FQHS
measured in the 30-nm-wide QW sample. It increases monotonically with
increasing $B$, consistent with our expectation that the 7/3 state
should become stronger when $E_F$ moves from the S1$\uparrow$ to the
A0$\uparrow$ level \footnote{Note that we expect the S1$uparrow$ and
  A0$uparrow$ levels to cross at $\nu=7/3$ at slightly larger $B$
  ($\sim 0.1$ T) compared to $\nu=5/2$ because of the density
  dependence of $\Delta$.}. We do indeed observe a strong rise in
$^{7/3}\Delta$ at this field. Note also in Fig. 4(e) that, at the
lowest fields and far from the crossing, $^{5/2}\Delta$ and
$^{7/3}\Delta$ in the 30-nm sample are of very similar magnitude, as
seen previously is standard single-subband 2DESs when $E_F$ lies in
the S1$\uparrow$ level. However, it appears from Fig. 4(e) data that
$^{7/3}\Delta$ much exceeds $^{5/2}\Delta$ even before the crossing
occurs.

In conclusion, we studied the stability of the $\nu$ = 5/2 FQHS when
the lowest LL of the antisymmetric electric subband (A0$\uparrow$)
crosses the second LL of the symmetric subband (S1$\uparrow$). The
5/2 FQHS is remarkably robust when $E_F$ lies in the S1$\uparrow$ LL
even as the A0$\uparrow$ level is brought to within $\sim$ 1 K of the
S1$\uparrow$ level. As the crossing is reached the 5/2 state
abruptly disappears, suggesting a first-order transition from a FQHS
to a metallic state.

\begin{acknowledgments}
  We acknowledge support through the Moore Foundation and the NSF
  (DMR-0904117 and MRSEC DMR-0819860) for sample fabrication and
  characterization, and the DOE BES (DE-FG0200-ER45841) for
  measurements. A portion of this work was performed at the National
  High Magnetic Field Laboratory, which is supported by NSF
  Cooperative Agreement No. DMR-0654118, by the State of Florida, and
  by the DOE. We thank J. K. Jain, Z. Papic, and J. Shabani for
  illuminating discussions, and E. Palm, J. H. Park, T. P. Murphy and
  G. E. Jones for technical assistance.
\end{acknowledgments}

\bibliography{paper_v2}

\end{document}